\documentclass[preprint]{aastex}

\newcommand \mum{$\mu$m}

\newcommand \msol{M$_{\odot}$}

\newfont{\rten}{cmr10}

\begin{document}

\slugcomment{To appear in AJ: final version 14 Feb. 2003}

\normalsize

\title{Dynamical models of Kuiper Belt dust in the inner and outer Solar System}

\author{Amaya Moro-Mart\'{\i}n\altaffilmark{1} and  Renu Malhotra\altaffilmark{2}}

\email{amaya@as.arizona.edu; renu@lpl.arizona.edu} 

\altaffiltext{1}{Steward Observatory, University of Arizona,
933 N. Cherry Ave., Tucson, AZ 85721, USA}

\altaffiltext{2}{Department of Planetary Sciences, University of Arizona,
1629 E. University Boulevard, Tucson, AZ 85721, USA}

\begin{abstract}

We report several results related to the dynamical evolution of dust produced
in the Kuiper Belt (KB).  We show that its particle size frequency distribution 
in space is greatly changed from the distribution at production, as a results of the
combined effects of radiation forces and the perturbations of the planets.  
We estimate the contribution of KB dust to the zodiacal cloud by calculating 
the radial profile of its number density near the ecliptic.  We also study the 
contribution of KB dust to the population of interplanetary dust particles (IDPs) 
collected at Earth, by 
calculating geocentric encounter velocities and capture rates.
Our models show, in contrast with previous studies, that KB dust grains on 
Earth-crossing orbits have high eccentricities and inclinations and, 
therefore, their encounter velocities are similar to those of cometary grains 
and not to asteroidal grains.  We estimate that at most 25\% in number 
of captured 
IDPs have cometary {\it or} KB origin; the KB contribution may be as low 
as 1\%--2\%. We present the velocity field of KB dust throughout the solar system; 
this, together with the number density radial profile, is potentially 
useful for planning spacecraft missions to the outer solar system.

\end{abstract}


\section{Introduction}
\label{intro}
Kuiper Belt Objects (KBOs) are icy bodies that lie in a disk 
beyond Neptune's orbit. It is estimated that there are about 
10$^5$ objects with diameters greater than 100 km in the 30--50 AU annulus 
and a total population roughly 3 orders
of magnitude larger than that of the asteroid belt (Jewitt \& Luu~\citeyear{jewi00}).
The outer limit of the belt is presently not well determined
but may be near 50 AU (Chiang \& Brown~\citeyear{chia99}; Allen, 
Bernstein \& Malhotra~\citeyear{alle01}). Stern (\citeyear{ster96}) suggested
that mutual collisions among KBOs can generate debris at a rate of 
(0.0095--3.2)$\times$10$^{11}$ g s$^{-1}$.
Using this estimate, Yamamoto \& Mukai (\citeyear{yama98}) calculated 
a dust production rate of 
(0.0086--2.9)$\times10^7$ g s$^{-1}$ in particles smaller than 10~$\mu$m. 
The impacts of interstellar dust on KBOs is also a significant source of 
interplanetary dust particles. 
Yamamoto \& Mukai (\citeyear{yama98}) estimated
that if there are $\sim$10$^{13}$ KBOs of radius $\ge$0.1 km, the 
total dust production rate for particles smaller than 10$\mu$m is
(0.37--2.4)$\times10^6$ g s$^{-1}$ if the objects have hard icy surfaces, or
(0.85--3.1)$\times10^7$ g s$^{-1}$ if the objects are covered with icy 
particles smaller than the interstellar grain impactors.
Detectors on board {\it Pioneer 10} and {\it 11} have indeed been able to detect
in situ dust in the outer solar system 
(see Landgraf et al.~\citeyear{land02}). Interstellar dust grains 
are estimated to contribute less than one percent to the measured flux, 
therefore this dust is thought to have a solar system origin. The dynamical
signatures indicate that the dust detected between Jupiter and Saturn is 
due to short period external Jupiter family comets and 
short period Oort cloud comets. The dust detected beyond 10 AU (outside 
Saturn's orbit) is most likely produced by KBOs. If so, then a  KB dust 
production rate of 2$\times$10$^{14}$ particles per second 
(for particles between 0.01 and 6 mm) is needed to explain the measured fluxes. 
Assuming a fragmentation power law for the size distribution,
this corresponds to a dust production rate of 
$\sim5\times$10$^{7}$ g s$^{-1}$, in agreement with 
the theoretical estimates above (Landgraf et al.~\citeyear{land02}).

The study of hypervelocity micrometeoroid craters on lunar material 
and on the panels of the {\it Long Duration Exposure Facility} showed that
Earth accretes about 3$\times$10$^{7}$ kg of 
interplanetary dust particles (IDPs) every year 
(Gr\"{u}n et al.~\citeyear{grun85}; Love \& Brownlee~\citeyear{love93}). 
Is the KB a significant source of these accreted IDPs?
Kortenkamp \& Dermott (\citeyear{kort98}, hereafter KD98) have calculated 
capture rates
for IDPs of asteroidal and cometary origins. Based on these rates, and
on the fact that the analysis of IDPs collected in the stratosphere shows a 
small diversity of chemical compositions (Flynn~\citeyear{flyn95}), they 
argue that the sources of IDPs are very limited and lie mainly in the 
asteroid belt, with less than 25\% having cometary origin. 
KB dust particles were, however, not considered in their study. 

The possibility that the KB may be a significant source of IDPs and the 
zodiacal cloud 
was first suggested by Liou, Zook \& Dermott (\citeyear{liou96},
hereafter LZD96). They found that
(1) about 20\% of KB dust particles are able to reach the inner solar 
system and (2) these particles have small eccentricities 
and inclinations (similar to asteroidal grains) when they cross the orbit 
of the Earth, enhancing their 
chances of being captured and of surviving atmospheric entry. 
The significance of these results is that, as they explain, 
asteroids are certainly an important source of IDPs, but they can accrete 
material from only as far as $\sim$4 AU and it is not clear that organic 
material at such distances can survive the T-Tauri wind of the young Sun.
KB dust grains, on the other hand, can bring in
unaltered primitive material from the outer solar system, 
so they could potentially be a source of the earliest organic 
material that gave rise to life on Earth.  Comets could also bring material
from the outer solar system, but as LZD96 argue, their high eccentricities
and inclinations cause the released dust particles to be in highly
eccentric and inclined orbits. This translates into high encounter velocities 
with Earth ($>$10 km/s), making it difficult for any organic material 
to survive atmospheric entry. In their paper, however, they
do not actually calculate capture rates and entry velocities for KB dust 
grains, giving only qualitative estimates. 

In this paper, we are going to follow numerically, from source to sink, 
the evolution of several hundred of dust particles from the KB
under the combined effects of solar gravity, solar radiation pressure, 
Poynting-Robertson (P-R) and solar wind
drag, and the gravitational forces of 7 planets (excluding Mercury and Pluto).
The parent bodies of the dust particles are assumed to be distributed in orbits
with semimajor axis between 35 and 50 AU, eccentricities such that the 
perihelion distances are between 35 and 50, and inclinations between 
0 and 17$^\circ$, in approximate accord with current estimates of the orbital 
distribution of KBOs (Malhotra et al.~\citeyear{malh00}).
We consider particles of diameter 3--115 $\mu$m (assuming $\rho$=1 g/cm$^{3}$; 
equivalently 1--40 $\mu$m for $\rho$=2.7 g/cm$^{3}$).
The different particle sizes are referred to in terms of their $\beta$ value,
which is the dimensionless ratio of 
the radiation pressure force and the gravitational force. 
For spherical grains and a solar type star, 
$\beta$=5.7 $\times$ 10$^{-5}$ Q$_{pr}$/($\rho${\it b}), where $\rho$ 
and {\it b} are the density and radius of the grain in cgs units 
(Burns, Lamy \& Soter~\citeyear{burn79}). 
The quantity Q$_{pr}$ is the radiation pressure coefficient, 
a function of the physical properties of the grain and the wavelength of
the incoming radiation; the value we use is an average integrated 
over the solar spectrum. 
The sinks of dust included in our numerical simulations are (1) 
ejection into unbound orbits, (2) accretion into the planets, and 
(3) orbital decay to less than 0.5 AU heliocentric distance.  
A detailed description of the models and the numerical algorithm used
to integrate the equations of motion 
is given in Moro-Mart\'{\i}n \& Malhotra (\citeyear{ama02}).

In $\S$2, we present the radial density profiles derived from our 
models and the KB dust production rate from Landgraf et al.~(\citeyear{land02});
from these, we estimate the contribution of the KB dust to 
the zodiacal cloud. Assuming steady state, this leads us to an 
estimate of the total mass in the KB dust disk.  
We also describe how the effects of radiation forces and the planets'
perturbations change the particle size distributions. 
In $\S$3, we address the question of the KB contribution 
to the collected IDPs on Earth by calculating geocentric 
encountering velocities and capture rates. 
In $\S$4, we present the velocity field of the KB dust in the inner 
and outer solar system.
In $\S$5, we evaluate the magnitude of the Lorentz force, which is not
included in our models.
Finally, $\S$6 summarizes our results.

\section{Radial Distribution and Mass of the KB Dust Disk}
\label{radial}
Based on our models and on estimates of KB dust production rates, 
we can calculate the number density of dust in the KB and its mass.
The simulations yield radial profiles of the number density of dust for 
various particle sizes; the production rates are used to get the 
normalization of these profiles.  For the production rates in the KB, 
we use the observationally based estimate by Landgraf et al. 
(\citeyear{land02}) of 2$\times$10$^{14}$ particles per second
of radius between 0.01 and 6 mm. For the size distribution, 
we use a fragmentation power law,
$\it {n(b)db=n_0b^{-q}db=n_0b^{-3.5}db}$,
where $\it{b}$ is the particle radius. (This corresponds to a generic 
grain mass distribution in collisional equilibrium; see, e.g.,  
Dohnanyi ~\citeyear{dohn69}.)
Then, assuming a bulk density $\rho$=1 g/cm$^{3}$,
we calculate the following dust production rates 
(in particles per second with the size bins in parentheses; the size 
corresponding to the particle's diameter):
4.5$\times$10$^{16}$ (2.1--4.2 $\mu$m; $\beta$=0.4), 8.0$\times$10$^{15}$ 
(4.2--8.5 $\mu$m; $\beta$=0.2), 1.4$\times$10$^{15}$ 
(8.5--17 $\mu$m; $\beta$=0.1), 2.5$\times$10$^{14}$ (17--33.9 $\mu$m; 
$\beta$=0.05) and 4.9$\times$10$^{12}$ (81.3--162.6 $\mu$m; $\beta$=0.01). 
Because of the power law distribution, the size bins are chosen in such a 
way that they all have the same width in logarithmic scale; for each size bin,
we have assigned a single $\beta$-value, as indicated (from Liou \& Zook 
\citeyear{liou99}).  

\subsection{Radial Distribution}
The radial profiles of the number density of particles within 1 AU of the ecliptic, 
based on our dynamical models and
normalized using the dust production rates and the size bins listed above, 
are shown in Figures 1{\it a} and 1{\it b}. 
In our models, the parent bodies are assumed to be distributed in orbits
with semimajor axis and perihelion distances between 35 and 50 AU.
Upon release, a dust particle has the same position and velocity as 
its parent body, but its orbital elements are different as a result of radiation 
pressure.  The latter effectively causes the particle to feel a Sun 
less massive by a factor (1-$\beta$). The larger the $\beta$, the 
more its orbit differs from its parent body's. 
After a particle leaves its parent body, P-R drag and solar wind drag tend 
to circularize and decrease the semimajor axis of its orbit,
forcing particles to slowly
drift in toward the central star (Burns, Lamy \& Soter~\citeyear{burn79}). 
Assuming that the dust particles are being produced constantly, 
this drif creates a dust disk of wide radial extent.
Figure 1{\it a} shows the radial distribution 
of particles of five different values of $\beta$
in a fictitious KB disk unperturbed by planets. 

The radial distribution changes in the presence of planets.  
We have modeled the gravitational effects of seven of the planets, Venus through Neptune.
Two effects play a major role in the quasi-steady state distribution of KB dust
that we obtain in our models (see Fig. 1{\it b}): 

\begin{enumerate}

\item
{\it Gravitational resonances.---} The journey of the dust particles 
toward the central star is temporarily interrupted by the trapping of particles 
in Mean Motion Resonances (MMRs), mainly with the outermost planet, Neptune.
The particles accumulate at certain semimajor 
axes, leading to the creation of structure in the disk; this explains
the ``bumps'' that appear between 35 and 50 AU.
The structure is more prominent for the smaller $\beta$-values because the resonance
trapping is more efficient when the drag force is small. 

\item
{\it Gravitational scattering.---}
Massive planets scatter and eject dust particles out of the 
planetary system, an effect that is independent of $\beta$. 
Scattering is responsible for the most striking difference between 
Figures 1{\it a} and 1{\it b}: for particles larger than the blow-out size 
($\beta$$\leqslant$0.5), the scattering of dust by the giant 
planets is able to extend the disk beyond the boundaries set by radiation 
effects alone. This has important consequences on the size distribution that 
will be explored below.

\end{enumerate}

In Figure 1{\it c}, we show the radial profile of the total number 
density of particles with radius between 1.4 and 10 $\mu$m, 
compared to the 
{\it Voyager 1} number density estimate, inferred from dust impact rates of 
1.4--10 $\mu$m particles from 30 to 51 AU (Gurnett et al.~\citeyear{gurn97}; 
Jewiit \& Luu~\citeyear{jewi00}). 
The radial profiles are obtained as follows:  First, 
for each choice of particle bulk density $\rho$,
we assign each $\beta$-value to a particle size bin (the size
being the particle's diameter).
For $\rho=1$ g/cm$^3$, we define the following size bins:
2.8--4.2 $\mu$m ($\beta$=0.4); 4.2--8.5 $\mu$m ($\beta$=0.2), 8.5--17 $\mu$m 
($\beta$=0.1) and 17--20 $\mu$m ($\beta$=0.05). For $\rho=2.7$ g/cm$^3$, 
the size bins are:
2.8--5.3 $\mu$m ($\beta$=0.1), 5.3--13.3 $\mu$m 
($\beta$=0.05) and 13.3--20 $\mu$m 
($\beta$=0.01). 
Next, we normalize the radial profile obtained from our numerical models 
for each of the values of $\beta$ (i.e., for the corresponding size bins) using the 
dust production rates from Landgraf et al.~(2002) and assuming the power law 
size distribution stated above ({\it q}=3.5). 
As a final step, we add the contribution from all the size bins to obtain 
the total number density radial distribution shown in Figure 1{\it c}.
The differences between the profiles for the two choices of $\rho$ arise 
from the fact that the bulk density affects the correspondence between the 
$\beta$-value and the particle size, and the size in turn affects the
estimate of the dust production rate because of the assumed power law 
in the size distribution. 
When comparing the modeled radial profiles with the 
{\it Voyager 1} estimate, one should keep in mind that there are uncertainties 
in
the dust production rates and in the 
index of the power law (both of which determine the normalization
factors of the models), as well as in the {\it Voyager 1} 
number density estimate 
(which is based on a few impact events and also has some uncertainty in the 
sizes of the particles detected). Nevertheless, 
the two modeled radial profiles are in good agreement with the {\it Voyager 1}
observations.
We cannot favor silicate over icy composition for 
KB particles based on this comparison.

\placefigure{f1}

\subsection{Size Distribution}
Radiation forces and planetary perturbations change the size distribution
of dust particles, as the particles spread out from their site of production
at rates that are dependent on their size.
Figure 2 shows these effects in plots of the cumulative size 
distribution at various heliocentric distances throughout the KB dust disk.
(The {\it cumulative} size distribution is calculated by integrating the 
{\it differential} size distribution obtained from our models in
the size bins described above for $\rho$=1 g/cm$^{3}$). 
As we mentioned, the initial differential size distribution at the time 
of dust production is assumed to be a power law with  
{\it q}=3.5; it is represented in Figure 2 
as the thick line of slope -2.5, with the distance between the squares 
indicating our particle size binning ``resolution''.  
The other lines represent the cumulative size distribution
obtained in our models at five different heliocentric distances: 
5, 21, 41, 81 and 141 AU, as indicated in the figure. 
Figure 2{\it a} shows results for a fictitious KB disk unperturbed 
by planets, while Figure 2{\it b} shows results for the KB dust disk perturbed
by the seven planets, Venus through Neptune. 

The main features are  the following:
\begin{enumerate}

\item
Radiation forces alone change the differential size distribution, from the
original power law with {\it q }= 3.5, to another power law of smaller index
(see Fig. 2{\it a}).
This is due to the fact that radiation pressure ``kicks out'' the smaller
particles preferentially and P-R drag spreads out the smaller particles
faster than the bigger ones.

\item
This shallower power law (with slope of $\sim$-1.5, corresponding to 
a differential power law index {\it q}$\sim$2.5) is maintained constant 
throughout the disk at distances smaller than the aphelion of 
the parent bodies (parallel dotted, dashed and solid lines in Fig. 2{\it a}).  
At larger distances, however, we start to encounter 
the disk boundaries set by radiation pressure, which depend on the 
particle sizes. This explains the steeper size distributions 
found at 81 and 141 AU (only the smaller particles reach those larger
distances). 

\item
In the presence of planets, the size distribution changes
greatly at distances larger than the aphelion of the parent bodies: 
compare the shallower slopes obtained at 141 AU and 81 AU in Figure 2{\it b},
with the steep slopes at the same distances in Figure 2{\it a}. 
Unlike radiation pressure, gravitational scattering by the giant planets can
send larger particles to these larger distances, 
effectively spreading all the dust widely. 
As the figure shows, the dust distribution is no longer described
by a power law with a single index.

\item
The trapping of particles in MMRs with Neptune (between 35 and 50 AU), 
and the fact that large particles are more easily trapped, 
explains why the slope of the size distribution becomes more shallow 
at 41 AU (slope about -1.5, {\it q}$\sim$2.5) than at 5 and 21 AU (slope about 
-1.9, {\it q}$\sim$2.9) 
(compare solid and dashed lines with dotted line in Fig. 2{\it b}).
\end{enumerate}

Although some of these effects are minor, the large change in the size 
distribution described in point 3 is very significant. 
It is clear that the detection of an exoplanetary 
dust disk of wide radial extent (a hundred to thousands of AU)
does not necessarily imply the presence of dust-producing planetesimals 
at such large distances: gravitational scattering by giant planets 
can spread the dust to distances much larger than the aphelion of the parent 
bodies. The obvious question is whether this effect could be used 
to unambiguously infer the presence of giant planets. We plan to address this 
question in the future by studying the effect that the change in the particle 
size distribution has on the disk's spectral energy distribution.
\placefigure{f2}

\subsection{Total Dust Mass}
From our models, we estimate
the total mass of the KB dust disk to be,
$m_{KB~dust}\sim$1.2$\times$10$^{-11}$ \msol~ for particles with diameter
2.4--160 $\mu$m (assuming $\rho$=1 g/cm$^{3}$), or 
$m_{KB~dust}\sim$4.2$\times$10$^{-11}$ \msol~ for particle with diameter
0.8--150 $\mu$m (for $\rho$=2.7 g/cm$^{3}$). 
Using COBE observations at 140 and 240 $\mu$m, Backman, Dasgupta \&
Stencel (\citeyear{back95}) 
set an upper limit for the total mass of dust in the KB of $\sim$ 
3$\times$10$^{-10}$~\msol. 
Jewitt \& Luu (\citeyear{jewi00}) calculated that the mass
in particles with radius 1.4--10 $\mu$m to be
$\sim$4$\times$10$^{-14}$~\msol, 
based on a simple estimate using the {\it Voyager 1} number density;
the volume of an annulus with 30 AU inner radius, 50 AU outer radius, 
and 10 AU thickness; and the assumption of an average grain mass of 
2$\times$10$^{-14}$ kg. For this size range, and using the same 
size bins used for Figure 1{\it c},
our models predict a mass 
of 5.2$\times$10$^{-13}$~\msol~(for $\rho$=1.0 g/cm$^{3}$),
or 5.5$\times$10$^{-12}$~\msol~(for $\rho$=2.7 g/cm$^{3}$).

The uncertainties in the derived $m_{KB~dust}$ come not only from the 
dust production rates, but also from the fact that we are extrapolating 
the results from only five $\beta$-values to a wide range of particle sizes.
To estimate $m_{KB~dust}$, we do the following: (1) We count the number
of particles present in our five ``steady state'' models, each 
corresponding to a different $\beta$. The models assume an 
artificial dust production rate of 100 particles every 1000 years
(see Moro-Mart\'{\i}n \& Malhotra~\citeyear{ama02} for more details).
(2) We multiply the number of particles by the ratio of the 
dust production rates derived from Landgraf et al.~(\citeyear{land02}; in 
the size bins corresponding to the values of $\beta$ under consideration), to 
our artificial dust production rate. This gives us the total number
of particles in each of the five size bins. (3) To convert this number into 
mass, one must multiply by the particle mass. The particle mass that we 
attribute to each size bin is calculated using the 
fragmentation power law, so that the small particles have more weight 
because they are more abundant. (If we were to use the mass of the particle
that lies in the middle of the bin [corresponding to the modeled $\beta$], 
our total dust masses would be about 4.5 times larger). (4) Finally, we add 
together the masses from the five different size bins. This results in the 
values of $m_{KB~dust}$ quoted above.

\section{Is the KB a Significant Source of IDPs?}
\label{idp}

We have calculated Earth's capture rates and entry velocities for KB dust 
grains based upon our numerical models, and adopting the procedure
of KD98. We find that
(1) KB dust grains have higher eccentricities 
when crossing the orbit of the Earth than those found by LZD96 and 
(2) their encounter velocities and capture rates 
are more similar to dust grains of cometary origin than to asteroidal 
origin; this is contrary to the results of LZD96. 

We define a particle to be Earth crossing if its orbit overlaps that of Earth,
that is, 
$\it{q}<$$\it R$$<\it{Q}$, where $\it{q}$=$\it{a}$(1-$\it{e}$) is the perihelion 
of the particle, $\it{Q}$=$\it{a}$(1+$\it{e}$) is its aphelion and
{\it R} is the heliocentric distance of Earth,
0.9833 AU$<${\it R}$<$ 1.0167 AU. The encounter velocity
$\it{v}_{0}$  between the Earth and a particle 
on a crossing orbit was calculated following Kessler (\citeyear{kess81}). 
The effective capture cross-section $\sigma_{c}$ is given by 
$\sigma_{c}$=$\sigma_{\earth}$(1+$\it{{v}_{e}}^2$/$\it{{v}_{0}}^2$), where
$\it{v}_{e}$ is the escape velocity from the Earth (at an altitude of 
100 km, $\it{v}_{e}$=11.1 km s$^{-1}$) and $\sigma_{\earth}$ is Earth's geometric
cross section. The average spatial density at heliocentric distance {\it R} and
ecliptic latitude  $\it{l}$ of a population of dust particles with
orbital elements $\it{a}$, $\it{e}$ and $\it{i}$ is given by
\begin{equation}
{S(R,l)={1 \over 2{\pi^{3}}Ra[(\textrm{sin}^2i-\textrm{sin}^2l)(R-q)(Q-R)]^{1/2}}.}
\end{equation}
The fraction of this population captured by the Earth at position 
({\it R,l}) per unit time 
is {\it p}=$\it{v}_{0}$$\sigma_{c}${\it S}({\it R,l}). Following KD98,
for each of the Earth-crossing particles 
in our models we have calculated {\it S}({\it R,l}) at 360 positions along
Earth's orbit, with {\it R} and $\it{l}$ uniformly distributed in the range
0.9833 AU$<${\it R}$<$ 1.0167 AU  and -0$^{o}$.00035$<\it{l}<$ 0$^{o}$.00035.
Table 1 shows the results after averaging over these 360 positions and over 
the whole population of Earth-crossing particles. For comparison, the 
results from KD98 for asteroidal and cometary dust (with $\beta$=0.0469) 
are also included.

As Figure 3 shows, we find high eccentricities for KB dust grains, 
similar indeed to cometary dust and not to asteroidal dust,  
which implies a low spatial density and high encounter velocity, 
and, therefore, a low capture rate 
(see Table 1). The asteroidal dust particles, on the other 
hand, have lower eccentricities and inclinations, which translates into a 
higher capture rate. The discrepancies with LZD96 probably
arise from the different criterion used to identify particles on 
Earth-crossing orbits; LZD96's criterion, $\it{a}<$1, which is less 
precise than the one we adopt here, has a strong bias toward low-eccentricity 
orbits.

\placefigure{f3}

In order to estimate the relative contributions of various sources to the 
IDPs captured at Earth from 
the relative capture rates in Table 1 (which depends only on the
orbital elements of the population of Earth-crossing particles),
we need to know the relative contribution of each source to the 
number density of particles on Earth-crossing orbits. 
This problem has yet to be solved because the actual dust production rates 
from 
asteroids, comets and KBOs are highly uncertain 
and very model dependent. Since cometary and KB dust grains have similar 
capture rates, we can extend the results of KD98 to predict that cometary 
plus KB dust can represent 
more than half of the IDPs captured by Earth only if comets and KBOs 
together supply $\sim$95\% of the Earth-crossing particles. 
Based on modeling of the IRAS
dust bands, KD98 estimated that (5--25)\% of the Earth-crossing particles
originate in the asteroid families Eos, Themis and Koronis. 
Using Dermott et al.'s~(\citeyear{derm94})~1:3 
ratio of dust produced by asteroid
families to that produced by all asteroids, we then have that all asteroids
contribute (15--75)\%, leaving the rest for comets and KBOs.
In the extreme case that as much as 85\% has cometary or KBO origin, 
this suggests that because of the lower capture rates of these highly
eccentric grains, only 25\% of the collected IDPs will be supplied
by comets and KBOs. Interestingly, Brownlee, 
Joswiak \& Love (\citeyear{blee94}) concluded that, based upon the maximum 
temperature reached during atmospheric entry from the study of helium release, 
about 20\% of
IDPs $<$ 10 $\mu$m~ have entry velocities typical of cometary IDPs. This is in 
agreement with the above estimate. 
In the other extreme case, where 25\% of IDPs have cometary or KBO origin, 
our models, together with KD98's results, suggest that they will represent 
only about 2\% of the collected 
IDPs. Our conclusion from this exercise is that the KB can certainly 
be a source of IDPs but it is not as important as predicted by 
LZD96. 

The estimates above are for the relative contributions
from the different sources to the collected IDPs and depend on their 
relative contribution to the number density of particles on Earth-crossing 
orbits. We can also calculate the absolute contribution from the KB 
by using Landgraf et al.~(\citeyear{land02}) dust production rates 
and the capture rates in Table 1. Our models, together with the dust 
production rates, yield the number of particles on Earth-crossing orbits.
The capture rates are the fraction of this population that is captured by 
Earth every 10$^9$ yr. The multiplication of these two numbers leads to the following 
results: 
1.2$\times$10$^{5}$ kg yr$^{-1}$ (2.4--160 $\mu$m, 
$\rho$=1 g cm$^{-3}$), or 4.1$\times$10$^{5}$ kg yr$^{-1}$
(0.8--150 $\mu$m, $\rho$=2.7 g cm$^{-3}$). 
These numbers should be compared to the total mass influx of 
3$\times$10$^{7}$ kg yr$^{-1}$ inferred from the microcraters on the 
{\it Long Duration Exposure Facility} (Love \& Brownlee ~\citeyear{love93}).
The microcraters correspond to particles of radius between 2.5 and 250 $\mu$m, and
show a peak and a cutoff in the particle size distribution near 100 $\mu$m. 
The accreted KB dust mass represents between 0.4\% 
(assuming $\rho$=1 g cm$^{-3}$) and 1.4\% (assuming $\rho$=2.7 g cm$^{-3}$) 
of this total mass influx.
The same uncertainties in the dust mass estimates 
mentioned in $\S$2.3 apply here ---namely, if we were to use the mass of the 
particle that lies in the middle of each bin instead of weighting
the mass using the power law, the values would be 4.5 times 
larger. Also, because most of the mass is 
contained in the large particle sizes, these results depend on the maximum 
particle radius chosen. The conclusion, however, is clear: if 
Landgraf et al.~(\citeyear{land02}) KB dust production rates are correct,
then the KB presently provides only a few percent of the collected IDPs.

The delivery rate of KB dust to Earth's vicinity is calculated using the dust
production rates in $\S$2 and the percentage of particles that is able to drift
all the way into the Sun, which, as seen in Table 2, is also a function of $\beta$.
The delivery rates are (expressed in particles per second, 
followed by the bin sizes in parentheses) 
4.9$\times$10$^{15}$ (2.1--4.2 $\mu$m), 
1.2$\times$10$^{15}$ (4.2--8.5 $\mu$m), 
2.9$\times$10$^{14}$ (8.5--17 $\mu$m), 
4.7$\times$10$^{13}$ (17--33.9 $\mu$m) 
and 5.4$\times$10$^{11}$ (81.3--162.6 $\mu$m).
One should keep in mind, however, that these estimates, and the ones
in $\S\S$ 2 and 4, are rather 
model-dependent: the Landgraf et al dust production 
rate estimate makes assumptions about the KB parent bodies' orbits that 
are significantly different from the observed distribution; they also 
assume that there is no source of dust in the 10--30 AU region, and 
we have neglected the destruction of dust grains due to interstellar and 
mutual collisions.

\section{Velocities of KB Dust Grains}
\label{vel}
A study of the velocity field of KB dust is useful for predicting the flux of 
particles colliding with a spacecraft exploring the outer solar system 
(e.g. {\it New Horizons} and {\it Interstellar Probe}).\footnote{http://pluto.jhuapl.edu/ and http://interstellar.jpl.nasa.gov/}
This is of interest for planning dust detectors or dust analyzers,
as well as for estimating the potential hazard posed by dust collisions
to fast-moving spacecraft.
In order to provide some general estimates, we have used our models to 
calculate the noncircular velocity of the KB dust in the ecliptic: 
for each particle, 
the instantaneous circular velocity at that distance has been calculated
and has been subtracted from its actual velocity. The resulting
magnitude of noncircular velocities in the ecliptic presented in Figure 4 
corresponds to the average values of the particles that lie in 
square cells of 1 AU in size.
We find no significant azimuthal structure, except for the 
following: between 25 and 35 AU, the non-circular velocities show a small 
systematic azimuthal variation at the level of 10--20\%, 
with a maximum at Neptune's position, which may be due to the fact that 
the particles trapped in MMRs tend to avoid the planet in the resonance.
Figure 5 shows the radial profile of the ratio
between the non-ircular and the circular velocity averaged over azimuth.
The increase of the fractional noncircular velocity for heliocentric 
distances $r\gtrsim$50 AU
is due to the fact that only particles of large eccentricities are to be 
found at distances beyond the parent bodies.
The noncircular velocities tend to be higher for smaller particles 
(larger $\beta$), as expected from their larger eccentricities upon release.
\placefigure{f4}
\placefigure{f5}

\section{Other Physical Processes}
\label{other}
Our models do not include the effect of magnetic fields on charged
dust grains and the dust grain destruction processes (such as sublimation, 
sputtering and collisions). Below, we briefly comment on how this may affect 
the results presented here, but a comprehensive evaluation of these 
processes is beyond the scope of this paper.
\subsection{Effect of Heliospheric Magnetic Fields}
Dust grains are generally electrically charged, as a results of 
the ejection of 
photoelectrons and the accretion of ions and electrons. 
Inside the heliosphere, $\lesssim$ 150 AU, the grains are therefore 
subject to the Lorentz force exerted by the interplanetary magnetic field, 
while outside the heliosphere the interstellar magnetic field dominates.
The effects of solar wind magnetic forces on charged 
dust grains have been discussed previously (Parker~\citeyear{park64}; 
Consolmagno~\citeyear{cons79}; Morfill \& Gr\"{u}n~\citeyear{morf79}; 
Mukai~\citeyear{muka85}; 
Gustafson~\citeyear{gust94}; Fahr et al.~\citeyear{fahr95}; 
Gr\"{u}n \& Svestka~\citeyear{grun96}). 
Here we summarize the lines of argument that lead us to conclude that the 
omission of Lorentz forces in our modeling is not a significant limitation 
of our results.

The interplanetary magnetic field is known to have a complex structure 
and time behavior. The dipole component changes polarity every 11 years, 
with the 22 year solar cycle. Moreover, near the ecliptic these sign 
reversals 
take place more rapidly because of the presence of the heliospheric 
current sheet,
the extension of the Sun's magnetic equator into interplanetary space, 
separating regions of opposite polarity. 
At solar minimum, the current sheet extends from approximately 
$-25^\circ$ to $25^\circ$ from the solar equator. Particles within this 
latitude range cross the current sheet at least twice every solar rotation 
($\sim$27 days), or four or even six times if the current sheet is wrapped 
because of higher order terms in the magnetic field 
(Balogh~\citeyear{balo96}). 
At higher ecliptic latitudes, the particles cross the current sheet at 
least twice as they orbit the Sun.  
Therefore, the time-averaged effect of the Lorentz force will tend to 
vanish within 
a particle's orbital period, because the sign reversals are significantly 
faster than the orbital period of most KB particles.
(We note that 80--90\% of the KB dust grains are ejected by the giant 
planets [see Table 2], and therefore their orbital periods during their 
lifetimes are generally much larger than the 11.8 year period of the 
innermost giant planet, Jupiter.  However, we cannot rule out resonant effects 
for charged grains that remain in the vicinity of Jupiter and Saturn 
for extended periods of time, as they may be subject to Lorentz forces
of period comparable to their orbital periods.)

Parker (\citeyear{park64}) was the first to study the effect of this 
fluctuating interplanetary field on dust grains on non-inclined, circular 
orbits. Because the dominant component of the field is perpendicular 
to the radial solar wind vector, with a magnitude $\sim 3\times10^{-5}/{\it r}$(AU) 
gauss for heliocentric distances {\it r} exceeding a few AU 
(Parker~\citeyear{park63}, p. 138), he concluded that the Lorentz force
will scatter the grains out of the ecliptic plane, by perturbing the 
particle's inclinations while keeping the energy of the orbit unchanged. 
At the distance 
of the Earth, the scattering would be important only for grains 
$\lesssim$1 $\mu$m, for which the inclinations change significantly 
before P-R drag sweeps them into the Sun.
More recently, Fahr et al.~(\citeyear{fahr95}) estimated that 
the inclination change causes a negligible evolutionary effect on 
zodiacal dust particles $\gtrsim$10 $\mu$m. 
They found that for particles with inclinations 
$\it{i}$$\leqslant$15$^\circ$, where the bulk of the dust particles 
considered in this paper are, this effect is completely negligible 
compared to P-R migration rates because of the stochastic character of the 
electromagnetic force near the current sheet; 
for $\it{i}$$>$15$^\circ$ and circular or quasi-circular
orbits, the Lorentz force exactly cancels out when integrated over a complete 
orbit, whereas for more eccentric orbits, the orbit-averaged change in 
inclination turns out to be very small because the Lorentz force
reverses every 11 years with the solar cycle.

But as Parker (\citeyear{park64}) pointed out, in reality,
the interplanetary field also fluctuates in the direction
perpendicular to the ecliptic.  These fluctuations cause
a random walk in the semimajor axis of the particles.
Over a period of time $\Delta t$, the P-R effect will
dominate over Lorentz scattering provided that
$\langle\Delta a\rangle_{PR} \gg \langle\Delta a^2\rangle^{1/2}_{L}$.
Using Consolmagno's (\citeyear{cons79}) derivation for
$\langle\Delta a^2\rangle^{1/2}_{L}$ in a circular orbit,
Jokipii \& Coleman's (\citeyear{joki68}) estimates for
the fluctuating perpendicular component of the magnetic field
(based on measurements by {\it Mariner 4}),
Burns et al.'s (\citeyear{burn79}) expression for
$\langle\Delta a\rangle_{PR}$, and
adopting $q=bV/300$ esu for the particle's electric charge,
we can write the condition above as
${bQ_{pr}\over V} \gg 0.64\left({a^3\over \Delta t} \right)^{1/2}$,
where $b$ is the particle's radius in \mum, {\it V} is its electrical 
potential in
volts, $\it{a}$ is in AU, and $\Delta t$ is in years.
The dependence on the time $\Delta t$ arises from the fact that
the P-R effect causes a systematic drift in {\it a} that is 
proportional to $\Delta t$, while the fluctuating Lorentz force 
causes a diffusion in {\it a} that is proportional to 
$(\Delta t)^{1/2}$.  Scaling the comparison time $\Delta t$
by the orbital period of the particle, that is, 
$\Delta t = (ka)^{3/2}$, where $k$ is a numerical factor,
we find that the P-R effect will dominate Lorentz scattering for 
particle sizes $b \gg 3.2 \Big({a/1\hbox{AU}\over
k}\Big)^{3/4}\Big({V\over5\hbox{volt}}\Big)\Big({1\over
Q_{pr}}\Big)\mu\hbox{m}$.
Thus, for particles of radius larger than a few microns, the 
systematic P-R drift will exceed the random Lorentz scattering
on timescales from a few orbital periods in the inner solar system
to a few tens of orbital periods in the outer solar system.
Over the characteristic P-R drift timescale, $(a/\dot a)_{PR}$,
Lorentz scattering is negligible for the particle sizes and 
heliocentric distances in our models. We therefore consider 
that neglecting the Lorentz force does not constitute a major 
limitation of this work.

\subsection{Sublimation} 
Silicate grains can survive sublimation to distances 
less than $\sim\!0.5$ AU, whereas pure water-ice grains are unlikely to 
survive interior to $\sim\!4$ AU  (see estimates in $\S$6.2 of 
Moro-Mart\'{\i}n \& Malhotra~\citeyear{ama02}). 
For example, the staying time for a grain between 1 AU and 2 AU from the 
Sun on a 
orbit with $\it{a}$=10 AU and $\it{e}$=0.9 (i.e. {\it q}=1AU and 
{\it Q}=19AU) 
is $\sim$10$^7$ s, while the lifetime of an icy grain with a radius of 
163 \mum~is $\sim$10$^3$ s at 2 AU from the 
Sun (Mukai~\citeyear{muka86}; Gustafson~\citeyear{gust94}). Therefore, 
the icy grain cannot survive near the Earth due to quick sublimation.
KB grains are likely a mixture of silicates and ices.  While the ice fraction 
will sublimate quickly, the silicate remnant will likely survive to sub-Earth
perihelion distances.
Qualitatively, and for the size ranges considered in this paper, 
we expect that the rapid 
loss of the ice component will cause the grain's orbit to become more 
eccentric, as a result of the increased magnitude of radiation pressure on smaller 
grain sizes. Thus, our dynamical models would underestimate the eccentricities 
of KB grains on Earth-crossing orbits. (However, for smaller
grains of radii less than a few tenths of a micron, the effect would be the
opposite because $\beta$ decreases as the grain's radius decreases.) 
Furthermore, taking into account the sublimation of the icy fraction, 
our conclusion from $\S$3, that $\sim\!1\%$ of silicate IDPs may be 
from the KB, becomes an upper limit. The overall conclusion is still the 
same: most of the captured IDPs do not come from the KB.

\subsection{Sputtering} 
Sputtering by solar wind particles may cause mass loss and erosion of dust
grains, as well as chemical alteration of their surfaces.
The erosion rate is quite uncertain in existing literature. 
Most estimates are based on the analysis of {\it Apollo} 
samples of lunar soils
and related computer simulations and bombardment experiments. 
Some of these estimates are as follows:
McDonnell \& Flavill (\citeyear{mcdo74}) and McDonnell et al. 
(\citeyear{mcdo77}), 
estimated an erosion rate of 0.043 $\AA$ yr$^{-1}$ and 0.43 $\AA$ yr$^{-1}$,
respectively, on the basis of He$^+$ bombardment experiments. 
A few years later, Flavill et al. (\citeyear{flav80}) estimated  
0.025--0.045 $\AA$yr$^{-1}$, while 
Kerridge (\citeyear{kerr91}) estimated 0.0024 $\AA$ yr$^{-1}$ based on 
analysis of Ar$^{36}$ retention efficiency for solar wind implantation 
and its measures in a lunar sputtered surface.  In another independent
study, Johnson \& Baragiola (\citeyear{john91}) estimated erosion rates of
0.1--0.2, 0.01--0.03 and 0.002--0.003 $\AA$ yr$^{-1}$, 
where the two lower estimates 
take into account the decrease of sputtering efficiency due to the 
sticking of sputtered material to neighboring grains and to 
micrometeorite vapor deposition, respectively. 
Evidently, the estimated erosion rates differ by up to a factor of 200
in these studies. Most recently, Mukai et al.~(\citeyear{muka01})~suggest 
a rate of 0.1--0.2 $\AA$ yr$^{-1}$.

Adopting an erosion rate of 0.2 $\AA$ yr$^{-1}$ at 1 AU, 
and taking into account that it scales with heliocentric distance
roughly as {\it r}$^{-2}$, we can estimate the mass loss experienced by our
modeled KB dust grains. Our dynamical studies of KB dust show that most of 
the particles spend most of their time at $\it{a}$$>$20 AU,
and that their typical lifetime is $\sim$10$^7$ yr
(Moro-Mart\'{\i}n \& Malhotra~\citeyear{ama02} Figs. 3 and 10).
Consider a typical particle that spends 10$^7$ yr at 20 AU from the Sun.
The fraction of mass loss is $\sim$ 50\% for a 3 \mum~ particle, and it
scales as {\it b}$^{-1}$, where $\it{b}$ is the particle radius.
(This is likely an upper limit because the particles usually get trapped
in exterior MMRs with Neptune at $\it{a}$$>$30AU.) 
Of course, one would need to take into account that as the particles 
drift in as a result of P-R drag, their erosion rate increases because
of increased
solar wind flux at smaller heliocentric distance. 
Our dynamical studies show that typically particles spend less than
$\sim$10$^6$ yr inside 20 AU. We estimate that a 3 \mum~grain
will be almost completely destroyed before reaching the inner solar system,
while a 10 \mum~grain will suffer little erosion. 
If the the erosion rate is 100 times smaller than our adopted value
(and within the present uncertainties), the mass loss would be negligible 
in both cases. We may therefore conservatively conclude that grains 
$>$10 \mum~do not suffer significant erosion due to corpuscular sputtering. 

Sputtering-induced chemical alteration of dust grain surfaces may also
reduce the mass loss. Corpuscular sputtering preferentially depletes 
the surface regions of volatiles, but also causes implantation of ions 
that can change the chemistry of the grain surface by producing mixing and 
molecular bonding between layers of dissimilar materials.  
This may explain why IDPs, thought to be Van der Waals-bonded aggregates, 
can lose icy mantles and remain sufficiently stable to survive atmospheric 
entry. A blackened, sputter-resistant, highly carbonized and refractory 
surface layer can be created from organic and volatile mantles 
(Johnson \& Lanzerotti~\citeyear{john86}, Johnson~\citeyear{john90}, 
Mukai et al.~\citeyear{muka01}). Once this layer is formed, 
the efficiency of erosion by corpuscular sputtering will be reduced.

Our conclusions above are consistent with the findings of Mukai \& Schwehm 
(\citeyear{muka81}) and 
Johnson (\citeyear{john90}) who conclude that at the distances at which 
sputtering is important, the erosion is relatively small under present 
solar wind conditions but chemical alterations may be significant.
  
\subsection{Collisions} 
In the optically thin limit of interest in the present work, mutual 
collisions of dust grains are not significant, but grain destruction
due to collisions with interstellar grains may be
(Liou \& Zook,~\citeyear{liou99}). 
In Moro-Mart\'{\i}n \& Malhotra (\citeyear{ama02}; $\S$6.1), we compared 
the collisional lifetimes estimated by LZD96 to the dynamical 
lifetimes derived from our models.  We concluded that collisions with 
interstellar grains are likely to be important for KB dust 
particles with diameters 
from 6 to 50 \mum: smaller particles survive because they drift in 
fast, and larger particles survive because they are not destroyed by a 
single impact. Interstellar grain collisions therefore may affect the 
particle size distributions presented in $\S$2.2. It would be useful to 
address this in detail in a future study.

\begin{enumerate}
\item[]
We note here that one of our long-term goals, as part of the {\it SIRTF} 
FEPS Legacy project (principal investigator 
M. Meyer),\footnote{http://feps.as.arizona.edu} is to study the effect of planets and radiation on the particle 
size distribution in exo-planetary systems.  Considering that there are 
large uncertainties 
in the solar wind corpuscular sputtering effects,
as well as the interstellar grain flux and size distribution for our own 
solar system, we think it is not well-justified to introduce in our numerical 
models the effects 
of sputtering and collisions for systems where the 
interstellar dust environment would likely be even less well known.
\end{enumerate}

\section{Summary and Conclusions}
\label{concl}

\begin{enumerate}

\item
We have estimated the radial distribution of KB dust from our
dynamical models and the KB dust production rate estimates from 
Landgraf et al.~(\citeyear{land02}). (We neglect dust physical destruction 
processes.)
We find that the presence of planets has a very important effect on the 
distribution of dust:
for particles larger than the blow-out size ($\beta\leqslant0.5$), 
the gravitational scattering of dust by the giant planets is able to extend 
the disk beyond the boundaries set by radiation effects alone. 
We also find that it has important consequences for the dust size-frequency 
distribution (see below).

\item
The observation of dust disks of wide radial extent, a hundred to 
thousands of AU,
does not necessarily imply the presence of dust-producing planetesimals 
at such large distances, because the gravitational scattering by giant planets 
at much smaller semimajor axes can cause the dust to spread to distances much 
larger than the aphelion of the dust parent bodies. 

\item
Radiation forces alone change the differential size distribution from the
(assumed) initial power law of index $q=3.5$ at production, 
to a shallower power law with $q\approx2.5$, 
valid at distances smaller than the aphelion of the parent bodies. 
No large particles are found at larger distances, and consequently the size 
distribution there is very steep. However, when we account for planetary
perturbations, the size distribution changes greatly at these large distances.
Overall, we conclude that the combination of radiation forces and planetary
perturbations causes the dust disk to spread out and the dust size frequency
distribution to flatten (Figs. 1 and 2).
In a future study, we plan to investigate the potential of the latter effect 
for the detection of planets in debris disks.

\item
We estimate the total mass of the KB dust disk to be
$m_{\rm KB~dust}\sim1.2\times10^{-11}$ \msol~
(2.4--160 $\mu$m, $\rho$=1 g cm$^{-3}$), or 
$m_{\rm KB~dust}\sim4.2\times10^{-11}$ \msol~
(0.8--150 $\mu$m, $\rho$=2.7 g cm$^{-3}$). These estimates are
consistent with other KB dust mass estimates found in the literature.

\item
We find in our dynamical models that KB dust grains near Earth have high 
eccentricities and inclinations similar to those of cometary grains and not 
asteroidal grains (Fig. 3).  
(Sublimation of the volatile fraction of these grains in the inner solar 
system is likely to increase their eccentricities further.)
As a consequence, they have encounter velocities and capture rates 
similar to cometary dust values; this is contrary to previous results
(Liou et al.~\citeyear{liou96}).

\item
We estimate, following Kortenkamp \& Dermott (\citeyear{kort98}), 
that at most 25\% of IDPs captured by Earth 
have cometary {\it or} KB origin. Furthermore, using
Landgraf et al.'s~(\citeyear{land02}) estimates of KB dust production rates, 
we find that the KB presently provides no more than a few percent of the 
collected IDPs.

\item
We have present the velocity field of KB dust grains in the inner and outer solar
system (Figs. 4 and 5). This is potentially useful for planning dust detectors
on future spacecraft missions, as well as for estimating the hazard to
space probes in the outer solar system.

\item
We estimate that the Lorentz forces due to the interplanetary magnetic field
within the heliosphere are likely negligible for the particle sizes
considered in this paper.
Mainly as a result of the rapid reversals in magnetic field polarity with the solar 
cycle, and the wrapped structure of the heliospheric current sheet,
the effect of the Lorentz force will tend to average out within a particle's 
orbit.

\item
Some physical destruction processes on KB dust grains may affect their 
dynamical evolution significantly, and detailed analysis in warranted in
future studies.  We estimate that the effect of rapid sublimation of the 
volatile component of KB dust grains is to increase their Earth encounter 
velocities and to reduce their relative abundance among captured IDPs.
The effects of sputtering by the solar wind are insignificant for grain sizes 
exceeding $\sim 10$\mum. Collisional destruction by interstellar grains likely 
modifies the size frequency distribution further, beyond the effects considered
in our dynamical models.

\end{enumerate}

\begin{center} {\it Acknowledgments} \end{center}
We thank Hal Levison for providing the SKEEL computer code 
and R. Jokipii and the anonymous referee for helpful discussions and 
comments.  A. M.-M. thanks the {\it SIRTF} Science Center and IPAC 
for providing access to their facilities during the completion of this work.
A. M.-M. is supported by NASA contract 1224768 administered by JPL. 
R. M. is supported by NASA grants NAG5-10343 and NAG5-11661.

\clearpage

\begin{deluxetable}{lccccc}
\tablewidth{0pc}
\tablecaption{EARTH-CROSSING DUST GRAINS}
\tablehead{
\colhead{Source}  &
\colhead{Average Capture} & \colhead{Geocentric Encountering} \\
\colhead{} & \colhead{Rate (Gyr$^{-1}$)} & \colhead{Velocity (km s$^{-1}$)}
}
\startdata
Kuiper Belt:& & \\
~~~$\beta$=0.01 & 10.9 & 13.4 \\
~~~$\beta$=0.05 & 10.2 & 13.3 \\
~~~$\beta$=0.1 & 14.5 & 12.1 \\
~~~$\beta$=0.2 & 14.7 & 12.4 \\
~~~$\beta$=0.4 & 9.3 & 18.0 \\
Asteroids\tablenotemark{a}:& & \\
~~~Eos & 100 & 5 \\
~~~Themis & 390 & 4 \\
~~~Koronis & 660 & 3 \\
~~~Other & 170 & 6 \\
Comets\tablenotemark{a} & & \\
~~~Trapped\tablenotemark{b} & 35 & 11 \\
~~~Nontrapped\tablenotemark{b} & 5 & 17 \\
\tablenotetext{~}{NOTE.--Earth orbital elements: a=1 AU, e=0.0167, i=0.00035$^{o}$}
\tablenotetext{a}{Approximate values from KD98 Fig.24, 25 (for $\beta$=0.0469 and a ratio of solar wind to P-R drag sw=0.3)}
\tablenotetext{b}{Previously trapped and non-trapped in MMR with Jupiter}
\enddata
\end{deluxetable}

\clearpage

\begin{deluxetable}{lccccc}
\tablewidth{0pc}
\tablecaption{FINAL FATE OF KUIPER BELT DUST GRAINS}
\tablehead{
\colhead{}  &
\colhead{$\beta$=0.01} & \colhead{$\beta$=0.05} & \colhead{$\beta$=0.1} & \colhead{$\beta$=0.2} & \colhead{$\beta$=0.4}}
\startdata
Ejected\tablenotemark{a} :& & & & & \\
~~~Jupiter &32 & 38 (45) & 44 (35) & 40 (35) & 20 (45) \\
~~~Saturn &37 & 28 (30) & 23 (40) & 31 (40) & 32 (35) \\
~~~Uranus &5 & 8 (0) & 6 (0)  & 6 (0) & 13 (0) \\
~~~Neptune &13 & 4 (0)  & 3 (5) & 8 (5) & 21 (5) \\
~~~None & \nodata& \nodata & \nodata & \nodata & 3 \\ 
Drift in &11 & 19 (25) & 21 (20) & 15 (20) & 11 (15) \\
Hit planet:& & & & & \\
~~~Jupiter &1 & 1 (0) & 1  (0)& \nodata & \nodata \\
~~~Saturn &1 & 2 (0) & 1 (0) &  \nodata & \nodata \\
~~~Uranus &\nodata& \nodata & 1 (0) &  \nodata & \nodata \\
\tablenotetext{~}{NOTE.--Listed as percentages; Liou et al. 1996 results appear in parentheses}
\tablenotetext{a}{Planet of last encounter}
\enddata
\end{deluxetable}
\clearpage


\figcaption[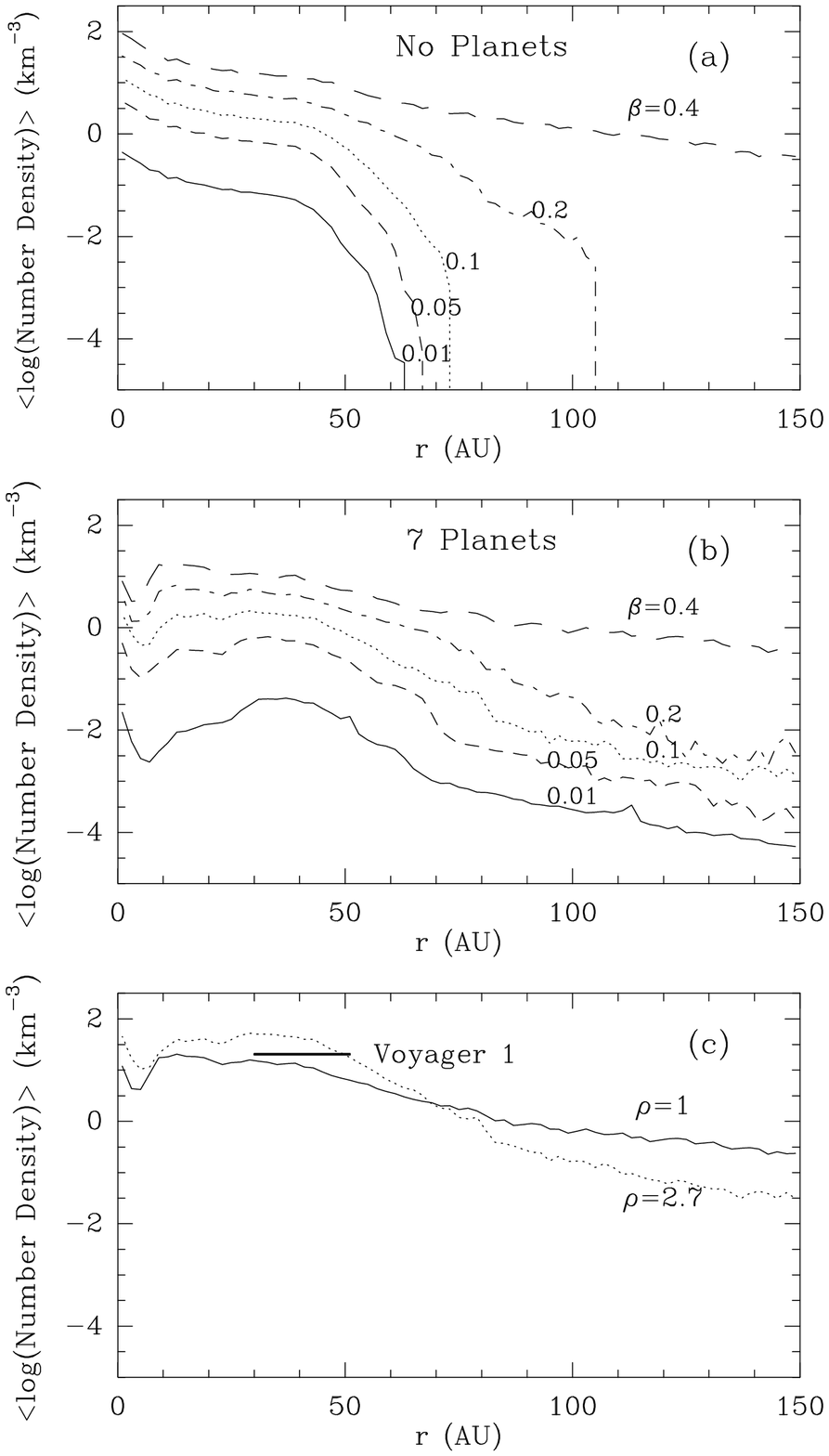]{\label{f1}
Radial profiles of the number density in the ecliptic for different 
values of $\beta$ (labeled). The normalization is calculated using the 
dust production rates in 
Landgraf et al.~(\citeyear{land02}), the size distribution
{\it n(b)db=n$_0b^{-3.5}$db} 
and the size bins in $\S$ 2.1. The total number of particles used to 
create these profiles are as follows (for $\beta$ of 0.01, 0.05, 0.1, 0.2 
and 0.4, respectively): with planets,  552,787, 180,832, 112,885, 61,234 and 
36,408; without planets: 1,107,811, 232,917, 125,398, 73,789 and 52,314.
({\it a}) A fictitious KB dust disk unperturbed by planets; 
({\it b}) KB dust disk perturbed by 7 planets;
({\it c}) comparison between the {\it Voyager 1} number density 
estimate ({\it thick solid line}; from Jewiit \& Luu~\citeyear{jewi00}) 
and a KB disk with 7 planets and two different particle bulk densities, 
for particle radius between 1.4 and 10 $\mu$m (see details in text).
}

\figcaption[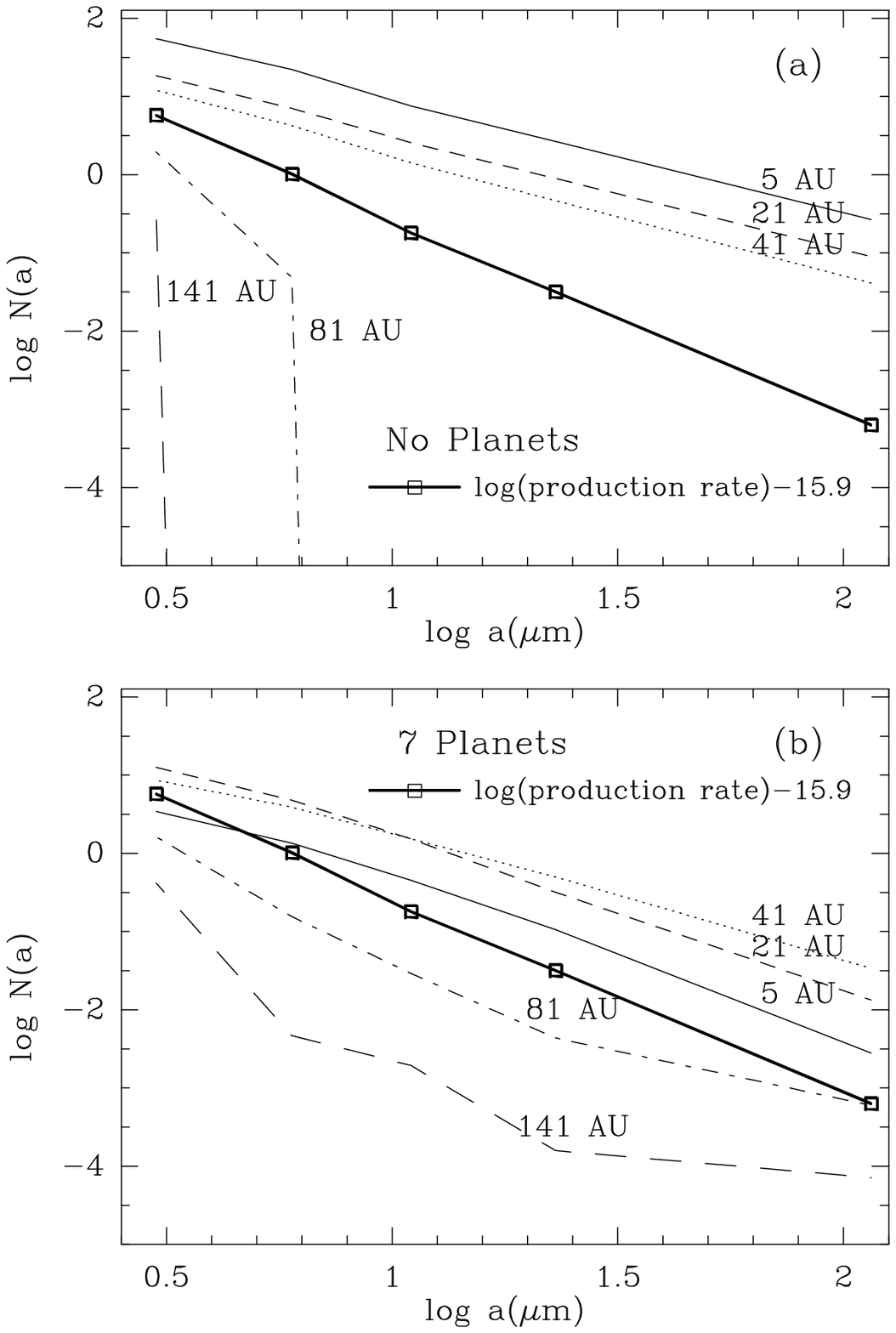]{\label{f2}
Effect of radiation forces and the presence of planets on the cumulative
size distribution (size bins in $\S$ 2.1). 
The thick solid line shows the size distributions at the time of 
dust production by the parent bodies
in units of number of particles per second. To fit in the figure, 
the line has been displaced by -15.9 dex. The slope of the cumulative 
distribution is -2.5, corresponding to a differential size distribution of
index -3.5.  The distance between the squares 
indicates the particle size binning ``resolution''.
The rest of lines the lines are the number density of particles in the 
ecliptic (in km$^{-3}$) at five different distances (indicated in the figure).
({\it a}) A fictitious KB dust disk unperturbed by planets; 
({\it b}) KB dust disk perturbed by 7 planets.
}

\figcaption[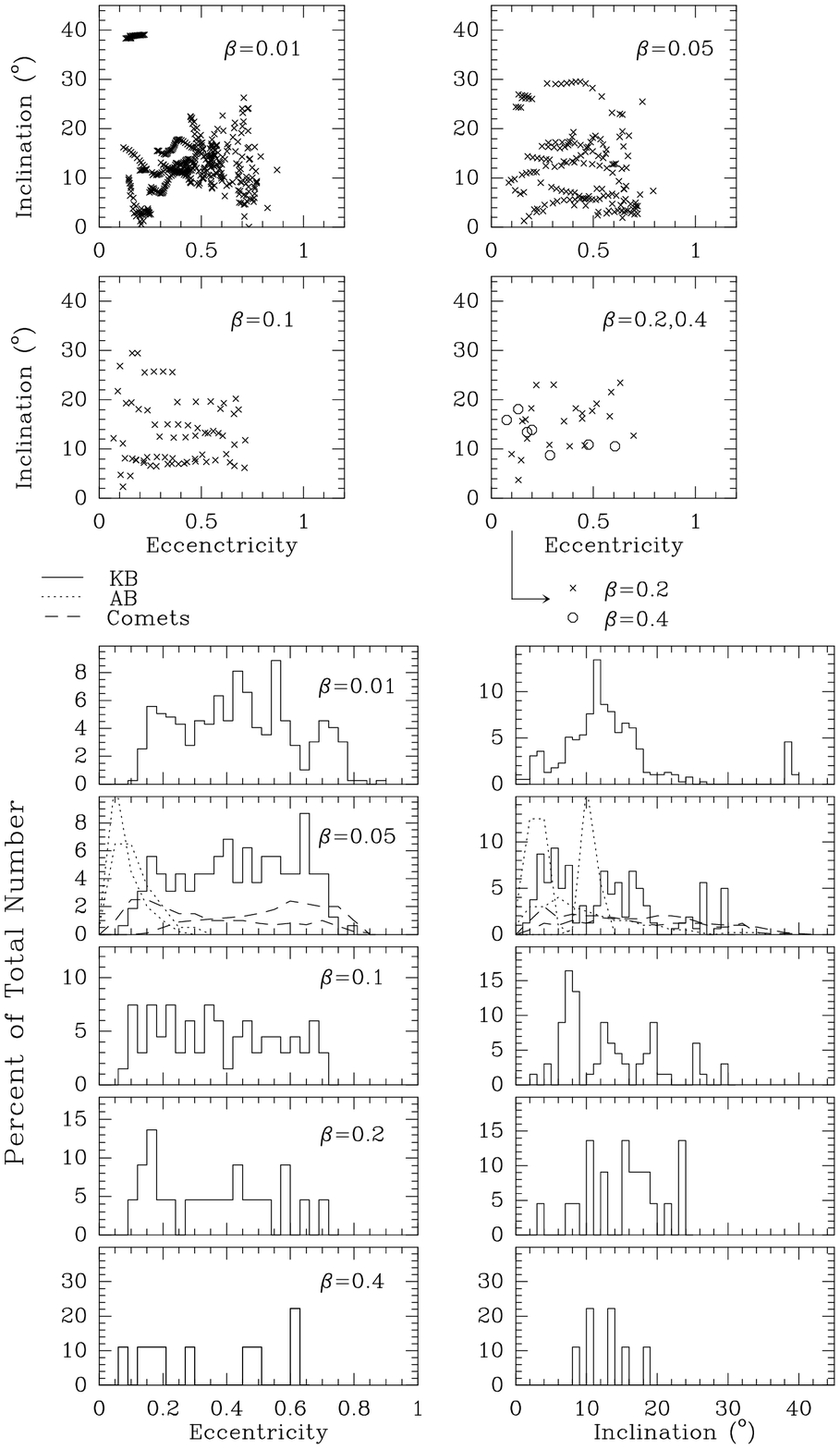]{\label{f3}
Eccentricities and inclinations of dust grains in Earth-crossing orbits.
For comparison, and in a different scale, the $\beta$=0.05 histograms 
show the distribution of eccentricities and inclination calculated by 
KD98 for dust particles with  $\beta$=0.0469, whose parent
bodies are the asteroid families Eos, Themis and Koronis, and  
the nonfamily asteroids ({\it dotted lines}) and the comets ({\it dashed lines}). 
The eccentricities and 
inclinations of the Earth-crossing KB dust grains are very different 
from those of the asteroidal dust, more resembling the distributions of cometary dust.}

\figcaption[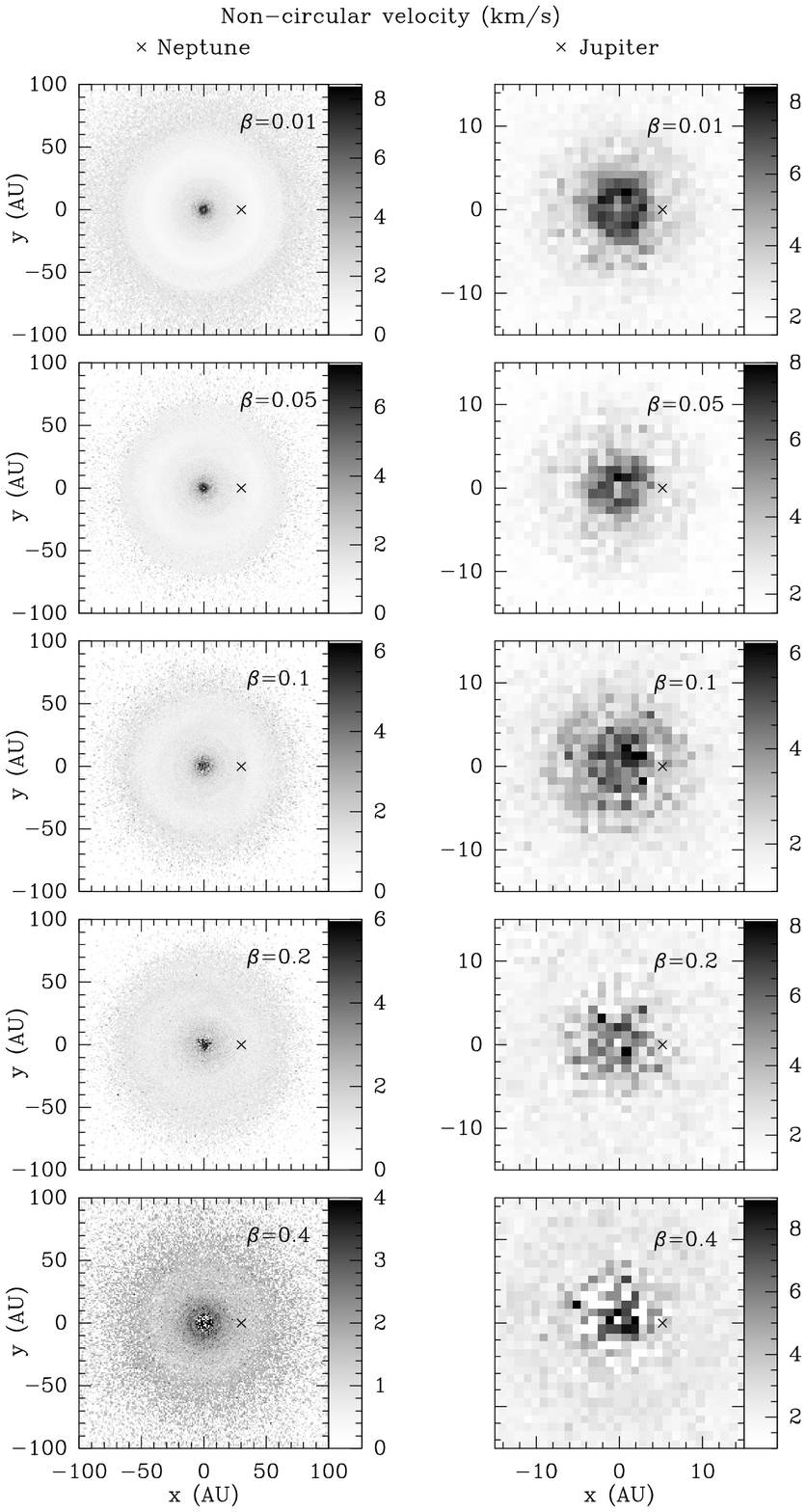]{\label{f4}
Non-circular velocity field (km s$^{-1}$) in the ecliptic for different particle 
sizes. The crosses indicate the position of Neptune ({\it left}) 
and Jupiter ({\it right}).}

\figcaption[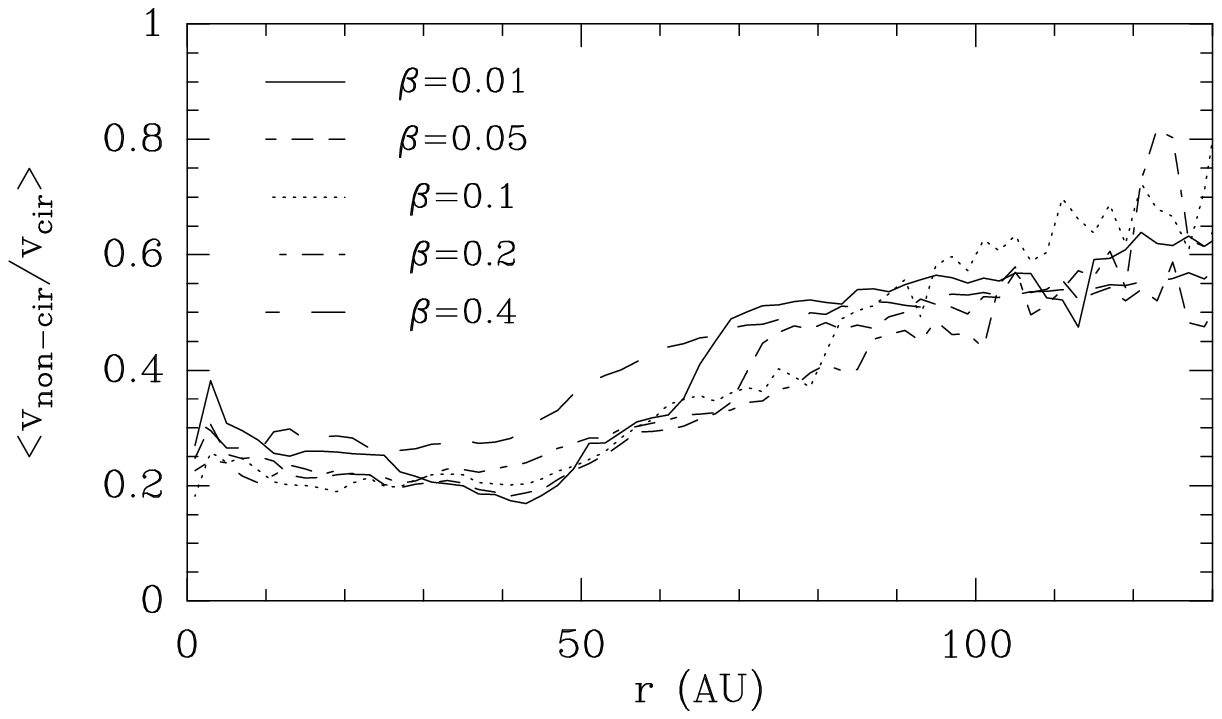]{\label{f5}
Radial profiles of the ratio between the non-circular and the circular 
velocity for different values of $\beta$.
}

\clearpage

\begin{figure}
\plotone{moromartin.fig1.ps}
\end{figure}

\begin{figure}
\plotone{moromartin.fig2.ps}
\end{figure}

\begin{figure}
\plotone{moromartin.fig3.ps}
\end{figure}

\begin{figure}
\plotone{moromartin.fig4.ps}
\end{figure}

\begin{figure}
\plotone{moromartin.fig5.ps}
\end{figure}

\clearpage

\end{document}